\documentstyle[fleqn,11pt]{book}
\begin{document}
\parindent 1.4cm
\large
\begin{center}
{\Large THE FEYNMAN-DE BROGLIE-BOHM PROPAGATOR FOR A SEMICLASSICAL
FORMULATION OF THE GROSS-PITAEVSKII EQUATION}
\end{center}
\begin{center}
{J. M. F. Bassalo$^{1}$,\ P. T. S. Alencar$^{2}$,\  D. G. da
Silva$^{3}$,\ A. Nassar$^{4}$\ and\ M. Cattani$^{5}$}
\end{center}
\begin{center}
{$^{1}$\ Funda\c{c}\~ao Minerva,\ Avenida Governador Jos\'e
Malcher\ 629 - CEP\ 66035-100,\ Bel\'em,\ Par\'a,\ Brasil /
E-mail:\ jmfbassalo@gmail.com}
\end{center}
\begin{center}
{$^{2}$\ Universidade Federal do Par\'a\ -\ CEP\ 66075-900,\
Guam\'a, Bel\'em,\ Par\'a,\ Brasil / E-mail:\ tarso@ufpa.br}
\end{center}
\begin{center}
{$^{3}$\ Escola Munguba do Jari, Vit\'oria do Jari\ -\ CEP\
68924-000,\ Amap\'a,\ Brasil / E-mail: \
danielgemaque@yahoo.com.br}
\end{center}
\begin{center}
{$^{4}$\ Extension Program-Department of Sciences, University of
California,\ Los Angeles, California 90024,\ USA / E-mail: \
nassar@ucla.edu}
\end{center}
\begin{center}
{$^{5}$\ Instituto de F\'{\i}sica da Universidade de S\~ao Paulo.
C. P. 66318, CEP\ 05315-970,\ S\~ao Paulo,\ SP, Brasil \ E-mail: \
mcattani@if.usp.br}
\end{center}
\par
Abstract:\ In this paper we present the Feynman-de Broglie-Bohm
propagator for a semiclassical formulation of the Gross-Pitaeviskii
equation.
\vspace{0.2cm}
\par
1.\ Introduction
\vspace{0.2cm}
\par
In the present work we investigate the Feynmam-de Broglie-Bohm
propagator for a semicassical formulation of the Gross-Pitaevskii
equation with the potencial $V(x,\ t)$ given by:
\begin{center}
{$V(x,\ t)\ =\ {\frac {1}{2}}\ m\ {\omega}^{2}(t)\ x^{2}$\ ,\ \ \ \ \ (1.1)}
\end{center}
which is the time dependent harmonic oscillator potential.
\par
2.\ Gross-Pitaeviskii Equation
\par
Em 1961[1,2], E. P. Gross and, independently, L. P. Pitaevskii proposed a
non-linear Schr\"{o}dinger equation to represent time dependent
physical systems, given by:
\begin{center}
{i\ ${\hbar}\ {\frac {{\partial}{\psi}(x,\ t)}{{\partial}t}}\ =\
-\ {\frac {{\hbar}^{2}}{2\ m}}\ {\frac {{\partial}^{2}\ {\psi}(x,\
t)}{{\partial}x^{2}}}\ +\ {\frac {1}{2}}\ m\ {\omega}^{2}(t)\
x^{2}\ {\psi}(x,\ t)\ +\ g{\mid}\ {\psi}(x,\ t)\ {\mid}^{2}\
{\psi} (x,\ t)$\ ,\ \ \ \ \ (2.1)}
\end{center}
where ${\psi}(x,\ t)$ is a wavefunction and $g$ is a constant.
\par
Writting the wavefunction ${\psi}(x,\ t)$ in the polar form, defined by
the Madelung-Bohm transformation[3,4], we get:
\begin{center}
{${\psi}(x,\ t)\ =\ {\phi}(x,\ t)\ e^{i\ S(x,\ t)}$\ ,\ \ \ \ \ (2.2)}
\end{center}
where $S(x\ ,t)$ is the classical action and ${\phi}(x,\ t)$ will be
defined in what follows.
\par
Substituting Eq.(2.2) into Eq.(2.1) and taking the real and imaginary
parts of the resulting equation, we get[5]:
\par
\begin{center}
{${\frac {{\partial}{\rho}}{{\partial}t}}\ +\ {\frac
{{\partial}({\rho}\ v_{qu})}{{\partial}x}}\ =\ 0$\ ,\ \ \ \ \ (2.3)}
\end{center}
\begin{center}
{${\hbar}\ {\frac {{\partial}S}{{\partial}t}}\ +\ {\frac {1}{2}}\ m\
v_{qu}^{2}(t)\ +\ {\frac {1}{2}}\ m\ {\omega}^{2}(t)\ x^{2}\ +\
V_{qu}\ +\ V_{GP}\ =\ 0$\ ,\ \ \ \ \ (2.4)}
\end{center}
\begin{center}
{${\frac {{\partial}v_{qu}}{{\partial}t}}\ +\ v_{qu}\ {\frac
{{\partial}v_{qu}}{{\partial}x}}\ +\ {\omega}^{2}(t)\ x\ =\ -\
{\frac {1}{m}}\ {\frac {{\partial}}{{\partial}x}}\ (V_{qu}\ +\ V_{GP})$\
,\ \ \ \ \ (2.5)}
\end{center}
where:
\begin{center}
{${\rho}(x,\ t)\ =\ {\phi}^{2}(x,\ t)$\ ,\ \ \ \ \ (2.6)\ \ \
(quantum mass density)}
\end{center}
\begin{center}
{$v_{qu}(x,\ t)\ =\ {\frac {{\hbar}}{m}}\ {\frac {{\partial}S(x,\
t)}{{\partial}x}}$\ ,\ \ \ \ \ (2.7)\ \ \ \ \ (quantum velocity)}
\end{center}
\begin{center}
{$V_{qu}(x,\ t)\ =\ -\ {\frac {{\hbar}^{2}}{2\ m}}\ {\frac {1}{{\sqrt
{{\rho}}}}}\ {\frac {{\partial}^{2}{\sqrt
{{\rho}}}}{{\partial}x^{2}}}\ =\ -\ {\frac {{\hbar}^{2}}{2\ m\
{\phi}}}\ {\frac {{\partial}^{2}{\phi}}{{\partial}x^{2}}}$\
,\ \ \ \ \ (2.8a,b)\ \ \ \ \ (Bohm quantum potential)}
\end{center}
and
\begin{center}
{$V_{GP}\ =\ g\ {\rho}$\ .\ \ \ \ \ (2.9)\ \ \ \ (Gross-Pitaevskii potential)}
\end{center}
\vspace{0.2cm}
\par
3.\ Feynman Propagator
\vspace{0.2cm}
\par
In 1948 [6], R. P. Feynman formulated the following
principle of minimum action for the quantum mechanics:
\begin{center}
{{\it The transition amplitude between the states ${\mid}\ a\ >$ and
${\mid}\ b\ >$ of a quantum-mechanical system is given by the sum of
the elementary contributions, one for each trajectory passing by
${\mid}\ a\ >$ at the time t$_{a}$ and by ${\mid}\ b\ >$ at the time
t$_{b}$. Each one of these contributions have the same modulus, but its
phase is the classical action S$_{c{\ell}}$ for each trajectory.}}
\end{center}
\par
This principle is represented by the following expression known as the
 "Feynman propagator":
\begin{center}
{$K(b,\ a)\ =\ {\int}_{a}^{b}\ e^{{\frac {i}{{\hbar}}}\ S_{c{\ell}}(b,\
a)}\ D\ x(t)$\ ,\ \ \ \ \ (3.1)}
\end{center}
with:
\begin{center}
{$S_{c{\ell}}(b,\ a)\ =\ {\int}_{t_{a}}^{t_{b}}\ L\ (x,\ {\dot {x}},\
t)\ dt$\ ,\ \ \ \ \ (3.2)}
\end{center}
where $L(x,\ {\dot {x}},\ t)$ is the Lagrangean and $D\ x(t)$ is the
Feynman's Measurement. It indicates that we must perform the integration
taking into account all the ways connecting the states ${\mid}\ a\ >$
and ${\mid}\ b\ >$.
\par
Note that the integral which defines $K(b,\ a)$\ is called "path
integral" or "Feynman integral" and that the Schr\"{o}dinger
wavefunction ${\Psi}(x,\ t)$ of any physical system is determined
using the expression (we indicate the initial position and initial
time by $x_{o}$ and $t_{o}$, respectively)[7]:
\begin{center}
{${\Psi}(x,\ t)\ =\ {\int}_{-\ {\infty}}^{+\ {\infty}}\ K(x,\ x_{o},\
t,\ t_{o})\ {\Psi}(x_{o},\ t_{o})\ dx_{o}$\ ,\ \ \ \ \ (3.3)}
\end{center}
with the quantum causality condition:
\begin{center}
{${\lim\limits_{t,\ t_{o}\ {\to}\ 0}}\ K(x,\ x_{o},\ t,\ t_{o})\ =\
{\delta}(x\ -\ x_{o})$\ .\ \ \ \ \ (3.4)}
\end{center}
\vspace{0.2cm}
\par
4.\ Calculation of the Feynman-de Broglie-Bohm Propagator for a
semiclassical formulation of the Gross-Pitaeviskii equation 
\vspace{0.2cm}
\par
The wavefunction ${\psi}(x,\ t)$ for the non-linear Gross-Pitaeviskii
is given by [8]:
\begin{center}
{${\Psi}(x,\ t)\ =\ [{\pi}\ {\sigma}(t)]^{-\ 1/4}\ exp\ {\Bigg
{[}}\ {\Big {(}}\ {\frac {i\ m\ {\dot {{\sigma}}}(t)}{4\ {\hbar}\
{\sigma}(t)}}\ -\ {\frac {1}{2\ {\sigma}(t)}}\ {\Big {)}}\ [x\ -\
q(t)]^{2}\ {\Bigg {]}}\ {\times}$}
\end{center}
\begin{center}
{${\times}\ exp\ {\Big {[}}\ {\frac {i\ m\ {\dot {q}}(t)}{{\hbar}}}\ [x\
-\ q(t)]\ +\ {\frac {i\ m\ v_{o}\ x_{o}}{{\hbar}}}\ {\Big {]}}\ {\times}$}
\end{center}
\begin{center}
{${\times}\ exp\ {\Bigg {[}}\ {\frac {i}{{\hbar}}}\
{\int}_{o}^{t}\ dt'\ {\Big {[}}\ {\frac {1}{2}}\ m\ {\dot
{q}}^{2}(t')\ -\ {\frac {1}{2}}\ m\ {\omega}^{2}(t')\ q^{2}(t')
-\ {\frac {{\hbar}^{2}}{2\ m\ {\sigma}(t')}}\ -\ g\ {\rho}(x,\ t')
{\Big {]}}\ {\Bigg {]}}$\ ,\ \ \ \ \ (4.1)}
\end{center}
where:
\begin{center}
{${\ddot {q}}\ +\ {\omega}^{2}(t)\ q\ =\ 0$\ ,\ \ \ \ \ (4.2)}
\end{center}
\begin{center}
{${\frac {{\ddot {{\sigma}}}(t)}{2\ {\sigma}(t)}}\ -\ {\frac
{{\dot {{\sigma}}}(t)^{2}}{4\ {\sigma}^{2}(t)}}\ +\
{\omega}(t)^{2}\ +\ {\frac {2\ g\ {\rho}(x,\ t)}{m\
{\sigma}(t)}}\ =\ {\frac {{\hbar}^{2}}{m^{2}\ {\sigma}^{2}(t)}}$\
,\ \ \ \ \ (4.3)}
\end{center}
\begin{center}
{${\rho}(x,\ t)\ =\ [{\pi}\ {\sigma}(t)]^{-\
1/2}\ e^{-\ {\frac {[x\ -\ q(t)]^{2}}{{\sigma}(t)}}}\ {\Big {(}}\ 1\
+\ {\frac {[x\ -\ q(t)]^{2}}{{\sigma}(t)}}\ {\Big {)}}$.\ \ \ \ \ (4.4)}
\end{center}
\par
with the following initial conditions are obeyed:
\begin{center}
{$q(0)\ =\ x_{o}\ ,\ \ \ {\dot {q}}(0)\ =\ v_{o}\ ,\ \ \ {\sigma}(0)\
=\ a_{o}\ ,\ \ \ {\dot {{\sigma}}}(0)\ =\ b_{o}$\ ,\ \ \ \ \ \
(4.5a-d)}
\end{center}
\par
Therefore, considering (4.1), the looked for Feynman-de
Broglie-Bohm propagator will be calculated using the expression
(3.3), in which we will put with no loss of generality, $t_{o}\ =\ 0$.
\par
Thus:
\begin{center}
{${\Psi}(x,\ t)\ =\ {\int}_{-\ {\infty}}^{+\ {\infty}}\ K(x,\ x_{o},\
t)\ {\Psi}(x_{o},\ 0)\ dx_{o}$\ .\ \ \ \ \ (4.6)}
\end{center}
\par
Initially let us define the normalized quantity:
\begin{center}
{${\Phi}(v_{o},\ x,\ t)\ =\ (2\ {\pi}\ a_{o}^{2})^{1/4}\ {\Psi}(v_{o},\
x,\ t)$\ ,\ \ \ \ \ (4.7)}
\end{center}
which satisfies the following completeness relation [9]:
\begin{center}
{${\int}_{-\ {\infty}}^{+\ {\infty}}\ dv_{o}\ {\Phi}^{*}(v_{o},\ x,\
t)\ {\Phi}(v_{o},\ x',\ t)\ =\ ({\frac {2\ {\pi}\ {\hbar}}{m}})\
{\delta}(x\ -\ x')$\ .\ \ \ \ \ (4.8)}
\end{center}
\par
Considering the eqs. (2.2), (2.6) and (4.7), we get:
\begin{center}
{${\Phi}^{*}(v_{o},\ x,\ t)\ {\Psi}(v_{o},\ x,\ t)\ =$}
\end{center}
\begin{center}
{$=\ (2\ {\pi}\ a_{o}^{2})^{1/4}\ {\Psi}^{*}(v_{o},\ x,\ t)\
{\Psi}(v_{o},\ x,\ t)\ =\ (2\ {\pi}\ a_{o}^{2})^{1/4}\ {\rho}(v_{o},\
x,\ t)\ \ \ {\to}$}
\end{center}
\begin{center}
{${\rho}(v_{o},\ x,\ t)\ =\ (2\ {\pi}\ a_{o}^{2})^{-\ 1/4}\
{\Phi}^{*}(v_{o},\ x,\ t)\ {\Psi}(v_{o},\ x,\ t)$\ .\ \ \ \ \
(4.9)}
\end{center}
\par
On the other side substituting the relation (4.9) in the
expression (2.3), integrating the result and using the
expressions (4.4) and (4.7) results [remembering that we have:\
${\Psi}^{*}\ {\Psi}({\pm}\ {\infty})\ \ \ {\to}\ \ \ 0$][7]:
\begin{center}
{${\frac {{\partial}({\Phi}^{*}\ {\Psi})}{{\partial}t}}\ +\ {\frac
{{\partial}({\Phi}^{*}\ {\Psi}\ v_{qu})}{{\partial}x}}\ =\ 0 \ \
\ {\to}$}
\end{center}
\begin{center}
{${\frac {{\partial}}{{\partial}t}}\ {\int}_{-\ {\infty}}^{+\
{\infty}}\ dx\ {\Phi}^{*}\ {\Psi}\ +\ {\int}_{-\ {\infty}}^{+\
{\infty}}\ {\frac {{\partial}({\Phi}^{*}\ {\Psi}\
v_{qu})}{{\partial}x}}\ dx\ =\ 0 \ \ \ {\to}$}
\end{center}
\begin{center}
{${\frac {{\partial}}{{\partial}t}}\ {\int}_{-\ {\infty}}^{+\ {\infty}}\
dx\ {\Phi}^{*}\ {\Psi}\ +\ ({\Phi}^{*}\ {\Psi}\ v_{qu}){\mid}_{-\
{\infty}}^{+\ {\infty}}\ =$}
\end{center}
\begin{center}
{$=\ {\frac {{\partial}}{{\partial}t}}\ {\int}_{-\ {\infty}}^{+\ {\infty}}\
dx\ {\Phi}^{*}\ {\Psi}\ +\ (2\ {\pi}\ a_{o}^{2})^{1/4}\ ({\Psi}^{*}\
{\Psi}\ v_{qu}){\mid}_{-\ {\infty}}^{+\ {\infty}}\ =\ 0\ \ \ {\to}$}
\end{center}
\begin{center}
{${\frac {{\partial}}{{\partial}t}}\ {\int}_{-\ {\infty}}^{+\
{\infty}}\ dx\ {\Phi}^{*}\ {\Psi}\ =\ 0$\ .\ \ \ \ \ (4.10)}
\end{center}
\par
The relation (4.10) shows that the integration is time
independent. Consequently:
\begin{center}
{${\int}_{-\ {\infty}}^{+\ {\infty}}\ dx'\ {\Phi}^{*}(v_{o},\ x',\ t)\
{\Psi}(x',\ t)\ =\ {\int}_{-\ {\infty}}^{+\ {\infty}}\ dx_{o}\
{\Phi}^{*}(v_{o},\ x_{o},\ 0)\ {\Psi}(x_{o},\ 0)$\ .\ \ \ \ \ (4.11)}
\end{center}
\par
Multiplying the relation shown in eq.(4.11) by ${\Phi}(v_{o},\
x,\ t)$ and integrating over $v_{o}$ and using the expression
(4.8), we will obtain [remembering that ${\int}_{-\
{\infty}}^{+\ {\infty}}\ dx'\ f(x')\ {\delta}(x' -\ x)\ = f(x)$]:
\begin{center}
{${\int}_{-\ {\infty}}^{+\ {\infty}}\ {\int}_{-\ {\infty}}^{+\
{\infty}}\ dv_{o}\ dx'\ {\Phi}(v_{o},\ x,\ t)\ {\Phi}^{*}(v_{o},\ x',\ t)\
{\Psi}(x',\ t)$\ =}
\end{center}
\begin{center}
{=\ ${\int}_{-\ {\infty}}^{+\ {\infty}}\ {\int}_{-\
{\infty}}^{+\ {\infty}}\ dv_{o}\ dx_{o}\ {\Phi}(v_{o},\ x,\ t)\
{\Phi}^{*}(v_{o},\ x_{o},\ 0)\ {\Psi}(x_{o},\ 0)\ \ \ {\to}$}
\end{center}
\begin{center}
{${\int}_{-\ {\infty}}^{+\ {\infty}}\ dx'\ ({\frac {2\ {\pi}\
{\hbar}}{m}})\ {\delta}(x'\ -\ x)\ {\Psi}(x',\ t)\ =\ ({\frac {2\ {\pi}\
{\hbar}}{m}})\ {\Psi}(x,\ t)$\ =}
\end{center}
\begin{center}
{=\ ${\int}_{-\ {\infty}}^{+\ {\infty}}\ {\int}_{-\ {\infty}}^{+\
{\infty}}\ dv_{o}\ dx_{o}\ {\Phi}(v_{o},\ x,\ t)\ {\Phi}^{*}(v_{o},\
x_{o},\ 0)\ {\Psi}(x_{o},\ 0)\ \ \ {\to}$}
\end{center}
\begin{center}
{${\Psi}(x,\ t)\ =\ {\int}_{-\ {\infty}}^{+\ {\infty}}\ {\Big {[}}\
({\frac {m}{2\ {\pi}\ {\hbar}}})\ {\int}_{-\ {\infty}}^{+\ {\infty}}\
dv_{o}\ {\Phi}(v_{o},\ x,\ t)\ {\times}$}
\end{center}
\begin{center}
{${\times}\ {\Phi}^{*}(v_{o},\ x_{o},\ 0)\ {\Big {]}}\ {\Psi}(x_{o},\
0)\ dx_{o}$\ .\ \ \ \ \ (4.12)}
\end{center}
\par
Comparing the relations (4.6) and (4.12), we have:
\begin{center}
{$K(x,\ x_{o},\ t)\ =\ {\frac {m}{2\ {\pi}\ {\hbar}}}\ {\int}_{-\
{\infty}}^{+\ {\infty}}\ dv_{o}\ {\Phi}(v_{o},\ x,\ t)\
{\Phi}^{*}(v_{o},\ x_{o},\ 0)$\ .\ \ \ \ \ (4.13)}
\end{center}
\par
Substituting the relations (4.1) and (4.7) in the equation
(4.13), we obtain the Feynman-de Broglie-Bohm Propagator for a
semiclassical formulation of the Gross-Pitaeviskii equation that we
were looking for, that is [remembering that ${\Phi}^{*}(v_{o},\ x_{o},\
0)\ =\ exp\ (-\ {\frac {i\ m\ v_{o}\ x_{o}}{{\hbar}}})$]:
\begin{center}
{$K(x,\ x_{o};\ t)\ =\ {\frac {m}{2\ {\pi}\ {\hbar}}}\ {\int}_{-\
{\infty}}^{+\ {\infty}}\ dv_{o}\ {\sqrt {{\frac
{a_{o}}{a(t)}}}}\ {\times}$}
\end{center}
\begin{center}
{${\times}\ exp\ {\Big {[}}\ {\Big {(}}\ {\frac {i\ m\ {\dot
{{\sigma}}}(t)}{2\ {\hbar}\ {\sigma}(t)}}\ -\ {\frac {1}{4\
{\sigma}^{2}(t)}}\ {\Big {)}}\ [x\ -\ q(t)]^{2}\ +\ {\frac {i\ m\ {\dot
{q}}(t)}{{\hbar}}}\ [x\ -\ q(t)]\ {\Big {]}}\ {\times}$}
\end{center}
\begin{center}
{${\times}\ exp\ {\Bigg {[}}\ {\frac {i}{{\hbar}}}\
{\int}_{o}^{t}\ dt'\ {\Big {[}}\ {\frac {1}{2}}\ m\ {\dot
{q}}^{2}(t')\ -\ {\frac {1}{2}}\ m\ {\omega}^{2}(t')\ q^{2}(t')
-\ {\frac {{\hbar}^{2}}{2\ m\ {\sigma}(t')}}\ -\ g\ {\rho}(x,\ t')
{\Big {]}}\ {\Bigg {]}}$\ .\ \ \ \ \ (4.14)}
\end{center}
where $q(t)$ and ${\sigma}(t)$ are solutions of the differential
equations given by the (4.2-4).
\par
In conclusion, we observe that the equations (4.1) and (4.14) we show
that when $g\ =\ 0$, then we obtains, respectively, the equations
(3.3.2.25) and (4.2.2.13) of the Reference [5], if ${\sigma}(t)\ =\ 2\
a^{2}(t)$,\ $q(t)\ =\ X(t)$\ and ${\frac {1}{2}}\ m\ {\omega}^{2}(t)\
q^{2}(t)\ =\ V[X(t)]$.
\newpage
\begin{center}
{NOTES AND REFERENCES}
\end{center}
\par
1.\ GROSS, E. P. 1961. Nuovo Cimento 20, 1766.
\par
2.\ PITAEVSKII, L. P. 1961. Soviet Physics (JETP) 13, 451.
\par
3.\ MADELUNG, E. 1926. Zeitschrift f\"{u}r Physik 40, 322.
\par
4.\ BOHM, D. 1952. Physical Review 85, 166.
\par
5.\ BASSALO, J. M. F., ALENCAR, P. T. S., CATTANI, M. S. D. e
NASSAR, A. B. 2003. T\'opicos da Mec\^anica Qu\^antica de de
Broglie-Bohm, EDUFPA.
\par
6.\ FEYNMAN, R. P. 1948. Reviews of Modern Physics 20, 367.\par
\par
7.\ FEYNMAN, R. P. and HIBBS, A. R. 1965. Quantum Mechanics and
Path Integrals, McGraw-Hill Book Company.
\par
8.\ BASSALO, J. M. F., ALENCAR, P. T. S., SILVA, D. G., NASSAR, A. B.
and CATTANI, M. arXiv:0910.5160v\ [quant-ph]\ 27\ October\ 2009.
\par
9.\ BERNSTEIN, I. B. 1985. Physical Review A 32, 1.
\par
10.\ BASSALO, J. M. F., ALENCAR, P. T. S., SILVA, D. G., NASSAR, A. B.
and CATTANI, M. arXiv:0902.3125\ [math-ph]\ 18\ February\ 2009.
\end{document}